\documentclass{PoS}

\usepackage{ifthen} 
\newboolean{uprightparticles}
\setboolean{uprightparticles}{false} %Set true for upright particle symbols
\usepackage{xspace} 
\usepackage{upgreek}
\usepackage{calc}       % for height measuring
\def\lhcb {\mbox{LHCb}\xspace}

\def\babar  {\mbox{BaBar}\xspace}
\def\belle  {\mbox{Belle}\xspace}

\def\dzero  {\mbox{D0}\xspace}

\def\MagUp {\mbox{\em Mag\kern -0.05em Up}\xspace}

\ifthenelse{\boolean{uprightparticles}}%
{

 \def\Pmu         {\ensuremath{\upmu}\xspace}                 
 \def\Pnu         {\ensuremath{\upnu}\xspace}                 
                  
 \def\Ppi         {\ensuremath{\uppi}\xspace}

 \def\Ptau        {\ensuremath{\uptau}\xspace}

 \def\PDelta      {\ensuremath{\Delta}\xspace}                 
 \def\PXi      {\ensuremath{\Xi}\xspace}                 
 \def\PLambda      {\ensuremath{\Lambda}\xspace}                 
 \def\PSigma      {\ensuremath{\Sigma}\xspace}                 
 \def\POmega      {\ensuremath{\Omega}\xspace}                 
 \def\PUpsilon      {\ensuremath{\Upsilon}\xspace}                 
 
 \def\PB      {\ensuremath{\mathrm{B}}\xspace}                 
                  
 \def\PD      {\ensuremath{\mathrm{D}}\xspace}

 \def\PK      {\ensuremath{\mathrm{K}}\xspace}

 \def\Pi      {\ensuremath{\mathrm{i}}\xspace}

 \def\Ps      {\ensuremath{\mathrm{s}}\xspace}

}
{

 \def\Pmu         {\ensuremath{\mu}\xspace}                 
 \def\Pnu         {\ensuremath{\nu}\xspace}                 
                  
 \def\Ppi         {\ensuremath{\pi}\xspace}

 \def\Ptau        {\ensuremath{\tau}\xspace}

 \mathchardef\PDelta="7101
 \mathchardef\PXi="7104
 \mathchardef\PLambda="7103
 \mathchardef\PSigma="7106
 \mathchardef\POmega="710A
 \mathchardef\PUpsilon="7107
                  
 \def\PB      {\ensuremath{B}\xspace}                 
                  
 \def\PD      {\ensuremath{D}\xspace}

 \def\PK      {\ensuremath{K}\xspace}

 \def\Pi      {\ensuremath{i}\xspace}

 \def\Ps      {\ensuremath{s}\xspace}

}

\makeatletter
\ifcase \@ptsize \relax% 10pt
  \newcommand{\miniscule}{\@setfontsize\miniscule{4}{5}}% \tiny: 5/6
\or% 11pt
  \newcommand{\miniscule}{\@setfontsize\miniscule{5}{6}}% \tiny: 6/7
\or% 12pt
  \newcommand{\miniscule}{\@setfontsize\miniscule{5}{6}}% \tiny: 6/7
\fi
\makeatother

\DeclareRobustCommand{\optbar}[1]{\shortstack{{\miniscule (\rule[.5ex]{1.25em}{.18mm})}
  \\ [-.7ex] $#1$}}

\def\mup        {{\ensuremath{\Pmu^+}}\xspace}
\def\mun        {{\ensuremath{\Pmu^-}}\xspace} % muon negative (\mum is taken)

\def\tauon      {{\ensuremath{\Ptau}}\xspace}

\def\taum       {{\ensuremath{\Ptau^-}}\xspace}

\def\neu        {{\ensuremath{\Pnu}}\xspace}
\def\neub       {{\ensuremath{\overline{\Pnu}}}\xspace}
\def\neum       {{\ensuremath{\neu_\mu}}\xspace}
\def\neumb      {{\ensuremath{\neub_\mu}}\xspace}
\def\neut       {{\ensuremath{\neu_\tau}}\xspace}
\def\neutb      {{\ensuremath{\neub_\tau}}\xspace}
\def\squark    {{\ensuremath{\Ps}}\xspace}

\def\pion   {{\ensuremath{\Ppi}}\xspace}

\def\pip    {{\ensuremath{\pion^+}}\xspace}
\def\pim    {{\ensuremath{\pion^-}}\xspace}

\def\pimp   {{\ensuremath{\pion^\mp}}\xspace}

\def\kaon    {{\ensuremath{\PK}}\xspace}
  \def\Kbar    {{\kern 0.2em\overline{\kern -0.2em \PK}{}}\xspace}

\def\KorKbar    {\kern 0.18em\optbar{\kern -0.18em K}{}\xspace}

\def\Kp      {{\ensuremath{\kaon^+}}\xspace}
\def\Km      {{\ensuremath{\kaon^-}}\xspace}
\def\Kpm     {{\ensuremath{\kaon^\pm}}\xspace}
\def\Kmp     {{\ensuremath{\kaon^\mp}}\xspace}

\def\Kstar   {{\ensuremath{\kaon^*}}\xspace}

  \def\Dbar    {{\kern 0.2em\overline{\kern -0.2em \PD}{}}\xspace}
\def\D       {{\ensuremath{\PD}}\xspace}

\def\DorDbar    {\kern 0.18em\optbar{\kern -0.18em D}{}\xspace}
\def\Dz      {{\ensuremath{\D^0}}\xspace}

\def\Dstar   {{\ensuremath{\D^*}}\xspace}

\def\Dstarp  {{\ensuremath{\D^{*+}}}\xspace}

\def\Dsm     {{\ensuremath{\D^-_\squark}}\xspace}

\def\Dsmp    {{\ensuremath{\D^{\mp}_\squark}}\xspace}

\def\B       {{\ensuremath{\PB}}\xspace}
\def\Bbar    {{\ensuremath{\kern 0.18em\overline{\kern -0.18em \PB}{}}}\xspace}

\def\BorBbar    {\kern 0.18em\optbar{\kern -0.18em B}{}\xspace}

\def\Bzb     {{\ensuremath{\Bbar{}^0}}\xspace}

\def\Bd      {{\ensuremath{\B^0}}\xspace}
\def\Bs      {{\ensuremath{\B^0_\squark}}\xspace}
\def\Bsb     {{\ensuremath{\Bbar{}^0_\squark}}\xspace}
\def\Bdb     {{\ensuremath{\Bbar{}^0}}\xspace}

  \def\Y#1S{\ensuremath{\PUpsilon{(#1S)}}\xspace}% no space before {...}!

\def\Lbar        {{\ensuremath{\kern 0.1em\overline{\kern -0.1em\PLambda}}}\xspace}
\def\LorLbar    {\kern 0.18em\optbar{\kern -0.18em \PLambda}{}\xspace}

\def\BF         {{\ensuremath{\cal B}}\xspace}

\def\BR         {\BF}
\def\to                 {\ensuremath{\rightarrow}\xspace}

\def\CP                {{\ensuremath{C\!P}}\xspace}

\def\AT#1     {\ensuremath{A_{\mathrm{T}}^{#1}}\xspace}           % 2

\def\C#1      {\ensuremath{\mathcal{C}_{#1}}\xspace}                       % 9
\def\Cp#1     {\ensuremath{\mathcal{C}_{#1}^{'}}\xspace}                    % 7
\def\Ceff#1   {\ensuremath{\mathcal{C}_{#1}^{\mathrm{(eff)}}}\xspace}        % 9  
\def\Cpeff#1  {\ensuremath{\mathcal{C}_{#1}^{'\mathrm{(eff)}}}\xspace}       % 7
\def\Ope#1    {\ensuremath{\mathcal{O}_{#1}}\xspace}                       % 2
\def\Opep#1   {\ensuremath{\mathcal{O}_{#1}^{'}}\xspace}                    % 7

\newcommand{\tev}{\ifthenelse{\boolean{inbibliography}}{\ensuremath{~T\kern -0.05em eV}\xspace}{\ensuremath{\mathrm{\,Te\kern -0.1em V}}}\xspace}
\newcommand{\gev}{\ensuremath{\mathrm{\,Ge\kern -0.1em V}}\xspace}
\newcommand{\mev}{\ensuremath{\mathrm{\,Me\kern -0.1em V}}\xspace}
\newcommand{\kev}{\ensuremath{\mathrm{\,ke\kern -0.1em V}}\xspace}
\newcommand{\ev}{\ensuremath{\mathrm{\,e\kern -0.1em V}}\xspace}
\newcommand{\gevc}{\ensuremath{{\mathrm{\,Ge\kern -0.1em V\!/}c}}\xspace}
\newcommand{\mevc}{\ensuremath{{\mathrm{\,Me\kern -0.1em V\!/}c}}\xspace}
\newcommand{\gevcc}{\ensuremath{{\mathrm{\,Ge\kern -0.1em V\!/}c^2}}\xspace}
\newcommand{\gevgevcccc}{\ensuremath{{\mathrm{\,Ge\kern -0.1em V^2\!/}c^4}}\xspace}
\newcommand{\mevcc}{\ensuremath{{\mathrm{\,Me\kern -0.1em V\!/}c^2}}\xspace}

\def\invfb   {\ensuremath{\mbox{\,fb}^{-1}}\xspace}

\def\gsim{{~\raise.15em\hbox{$>$}\kern-.85em
          \lower.35em\hbox{$\sim$}~}\xspace}
\def\lsim{{~\raise.15em\hbox{$<$}\kern-.85em
          \lower.35em\hbox{$\sim$}~}\xspace}

\def\tell1  {TELL1\xspace}
\def\ukl1   {UKL1\xspace}

\newcommand{\eg}{\mbox{\itshape e.g.}\xspace}

\def\RDstar {\ensuremath{\mathcal{R}(\Dstar) }\xspace }

\def\asld   {\ensuremath{a_{\rm sl}^{d}}\xspace}
\def\asls   {\ensuremath{a_{\rm sl}^{s}}\xspace}

\title{Search for new physics in semileptonic \B-decays}

\ShortTitle{Search for new physics in semileptonic \B-decays}

\author{\speaker{Suzanne Klaver}\thanks{On behalf of the \lhcb collaboration}\\
        The University of Manchester, Manchester, UK\\
        E-mail: \email{suzanne.klaver@cern.ch}}

\abstract{Semileptonic decays provide an excellent environment for testing the Standard Model (SM). Violation of lepton universality would be a smoking gun for physics beyond the SM. Using semi-tauonic \B decays, \lhcb finds a value of $\mathcal{R}(\Dstar) = \BR(\Bzb \to \Dstarp \taum \neutb)/\BR(\Bzb \to \Dstarp \mun \neumb) =  0.336 \pm 0.027 \rm{(stat)} \pm 0.030 \rm{(syst)}$, which is 2.1 standard deviations larger than the value expected from the SM. Moreover, the measurement of the \CP asymmetry in mixing of \Bs mesons is highly sensitive to physics beyond the SM. This article presents the latest result on semileptonic asymmetries; using the full Run 1 dataset, it is found that $\asls = (0.39 \pm 0.26 \rm{(stat)} \pm 0.20 \rm{(syst)})\%$, which is consistent with the Standard Model.}

\FullConference{38th International Conference on High Energy Physics\\
		3-10 August 2016\\
		Chicago, USA}

\begin{document}

\section{Introduction to semileptonic \B-decays}
Semileptonic decays are abundant and provide an excellent environment for precision physics as they can be precisely predicted theoretically. The outstanding vertex resolution of the \lhcb detector \cite{Alves:2008zz, LHCb-DP-2014-002} enables to carry out many exciting analyses involving semileptonic \B decays. This article summarises two results from \lhcb: a test of lepton universality in \RDstar and the measurement of \CP violation in \Bs-\Bsb mixing, both of which use the 3.0\invfb dataset recorded in 2011 and 2012.

\section{Testing lepton universality in \RDstar}
In the Standard Model, the branching fractions $\BR(\Bdb \to \Dstarp \taum \neutb)$ and $\BR(\Bdb \to \Dstarp \mun \neumb)$ differ only due to the difference in lepton mass between the two decays. However, in many extensions of the SM, enhanced couplings to the third generation are predicted, in particular models containing charged Higgs bosons. The measurement of the ratio of branching fractions:
\begin{equation}
\RDstar \equiv \frac{ \BR( \Bdb \to \Dstarp \taum \neutb ) } { \mathcal{B}( \Bdb \to \Dstarp \mun \neumb ) },
\label{eq:RDstar}
\end{equation}
therefore tests this idea of lepton universality present in the SM, which predicts a value of $\RDstar = 0.252 \pm 0.003$ \cite{Fajfer:2012vx}.

The \taum candidate is reconstructed as \taum \to \mun \neumb \neut, hence the signal channel has the same visible final-state as the control channel \Bdb \to \Dstarp \mun \neumb \cite{LHCb-PAPER-2015-025}. The selection for both decays is similar which combines a muon and the \Dstarp candidate that is reconstructed through $\Dstarp \to \Dz (\to \Km \pip) \pip$. In addition to the signal and normalisation channel, also several background processes are selected, which are suppressed by exploiting kinematic and topological properties.

\begin{figure}[h]\centering
\includegraphics[width=0.9\textwidth]{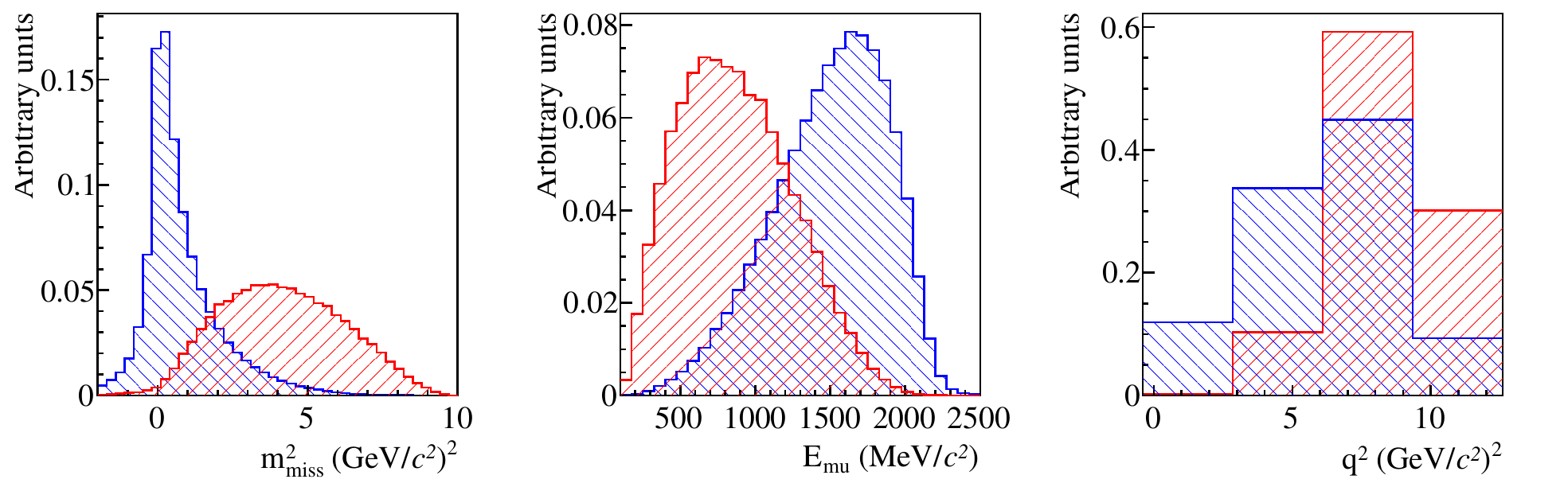}
\caption{\label{fig:RDst_MC} Simulated distributions of the kinematic variables that are used in the fits. In blue are the distributions from the normalisation channel: \Bzb \to \Dstarp \mun \neumb, and in red from the signal: \Bzb \to \Dstarp \taum \neutb. From left to right are the distributions of $\rm{m}_{\rm{miss}}^2$, $E_{\mu}^{*}$, and $q^2$.}
\end{figure}

A multidimensional fit is performed to measure to signal, normalisation and background components. The variables used for the fit are kinematic variables which are most discriminating between the signal and normalisation channel. These are the missing mass squared, $m_{\rm{miss}}^2$, the muon energy in the center-of-mass, $E_{\mu}^*$, and the squared four-momentum of the lepton system, $q^2$. Simulated distributions of the signal and normalisation channel of these variables are shown in Fig.~\ref{fig:RDst_MC}.

\begin{figure}
  \centering
  \setbox1=\hbox{\includegraphics[width=\textwidth]{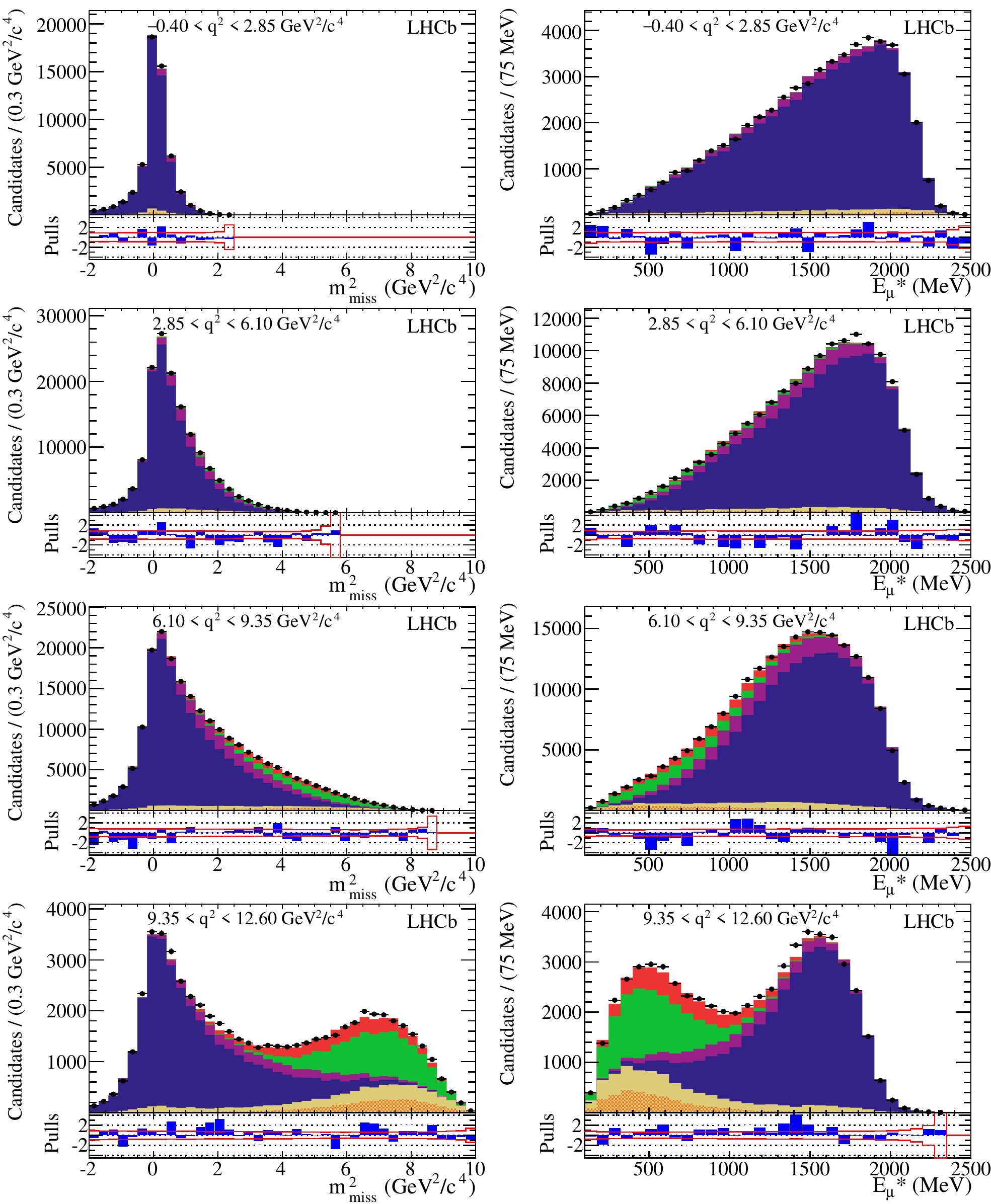}}  
  \includegraphics[width=\textwidth, trim={0 0 0 12.15cm},clip]{figs/RDst_Fig1.pdf}\llap{\makebox[10.5cm][l]{\raisebox{6.8cm}{\includegraphics[height=1.7cm]{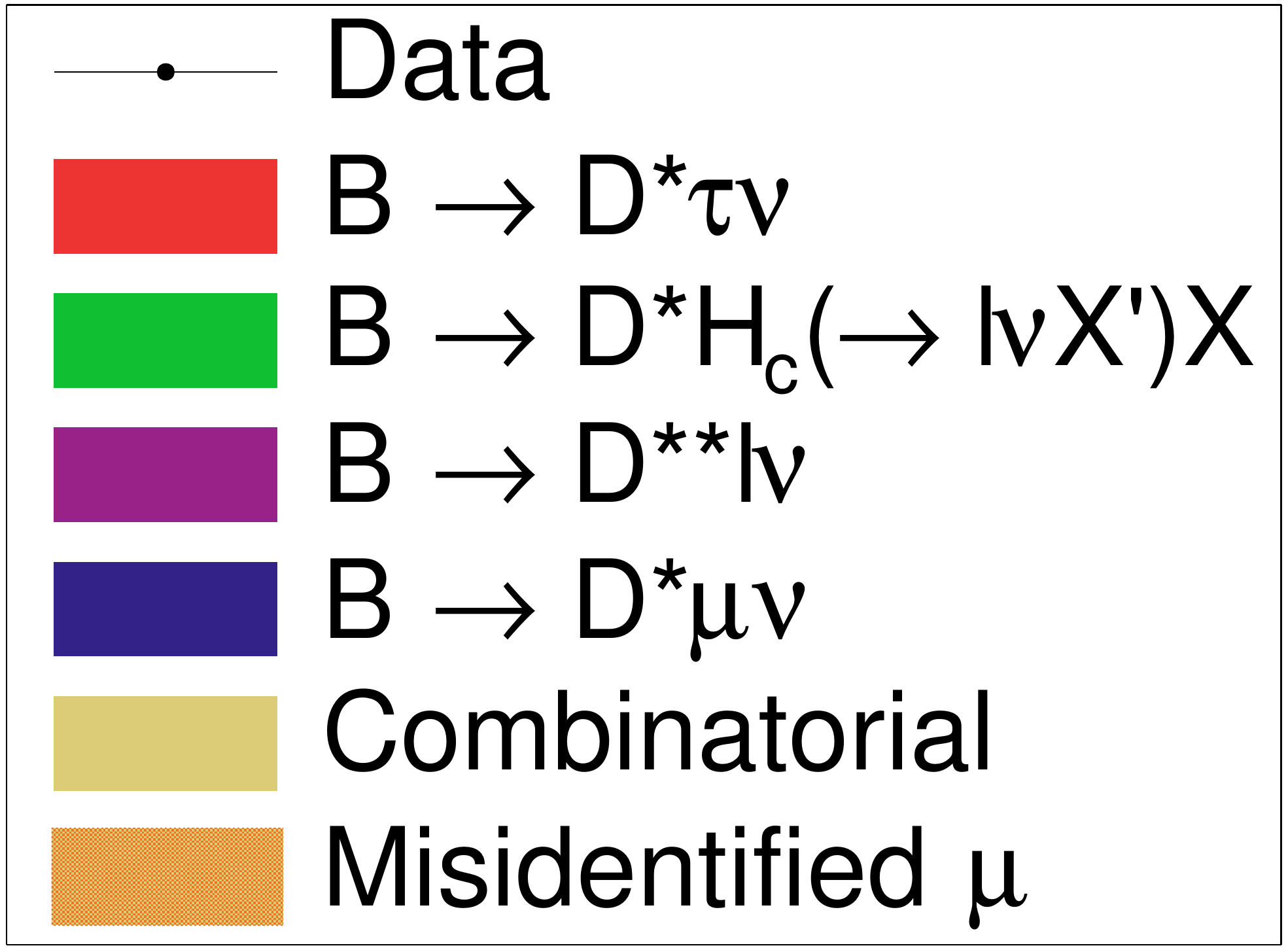}}}}
  \caption{\label{fig:RDst_fits} Distributions of $\rm{m}_{\rm{miss}}^2$ (left) and  $E_{\mu}^{*}$ (right) in the two highest $q^2$ bins of the signal data, overlaid with the projections of the fit model. The signal distributions are red, and the normalisation channel is blue.}
\end{figure}

The three-dimensional fit is performed using templates of the signal, normalisation, and background channels. Projections in $m_{\rm{miss}}^2$ and $E_{\mu}^*$ of the data and fits in the two highest bins of $q^2$ are shown in Fig.~\ref{fig:RDst_fits}. The shapes from the combinatorial and misidentified backgrounds are taken from data, while the physics background are modelled using control samples. The  \Bdb \to \Dstarp \taum \neutb is most visible in the highest $q^2$ bin.

\begin{figure}[h]\centering
\includegraphics[width=0.325\textwidth]{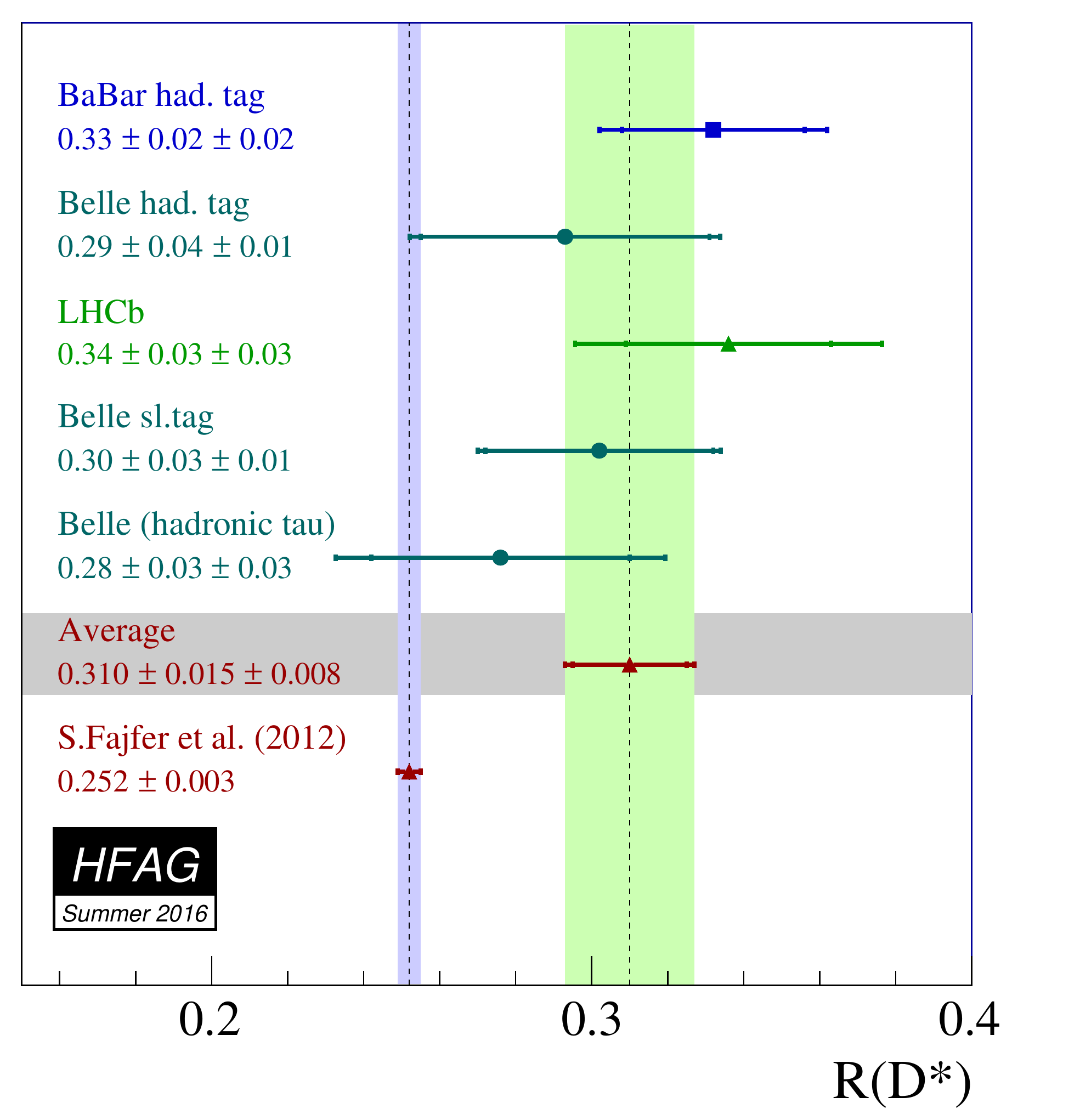}
\includegraphics[width=0.5\textwidth]{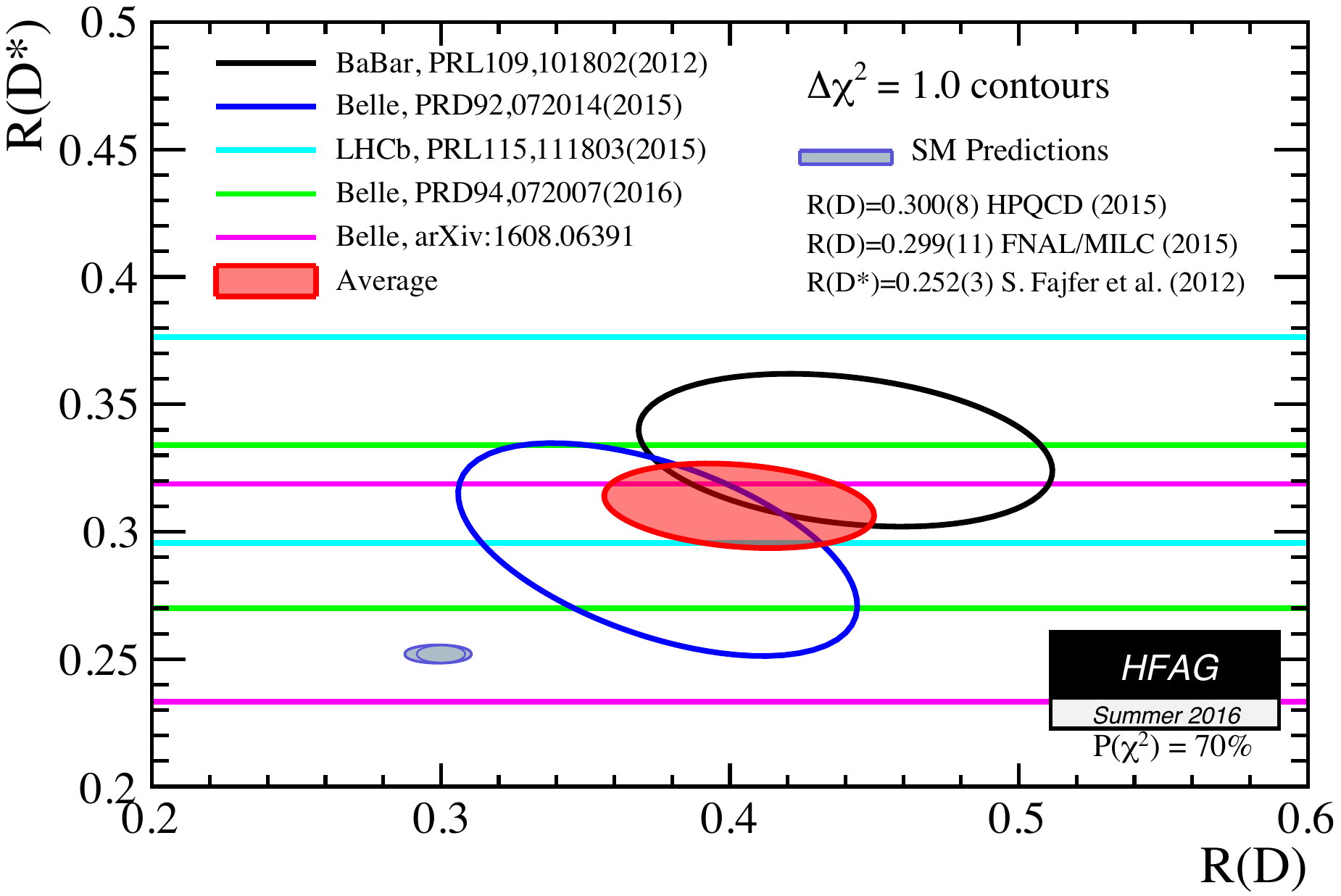}
\caption{\label{fig:RDst_RD} On the left: comparisons of the \lhcb $\mathcal{R}(\Dstar)$ result with those from \belle \cite{Bozek:2010xy,Huschle:2015rga,Abdesselam:2016cgx} and \babar \cite{Lees:2012xj}. On the right: the comparison of the \lhcb measurement of $\mathcal{R}(\Dstar)$ with the $\mathcal{R}(D)$ vs. $\mathcal{R}(\Dstar)$ measurements of \babar and \belle and the Standard Model predictions \cite{Fajfer:2012vx,Na:2015kha,Lattice:2015rga} from HFAG \cite{Amhis:2014hma}.}
\end{figure}

The \lhcb measurement is $\RDstar = 0.336 \pm 0.027 \rm{(stat)} \pm 0.030 \rm{(syst)}$, which is the first measurement of a \B to \tauon decay at a hadron collider. Fig.~\ref{fig:RDst_RD} (left) shows the \lhcb measurement of \RDstar and those performed by \belle \cite{Bozek:2010xy,Huschle:2015rga,Abdesselam:2016cgx} and \babar \cite{Lees:2012xj}, which indicates consistency between the experimental results. Fig.~\ref{fig:RDst_RD} (right) displays the $\mathcal{R}(D)$ vs. $\mathcal{R}(\Dstar)$ measurements and the current theoretical predictions, indicating a $3.9\sigma$ discrepancy between theory and experiment. This could be explained by \eg models containing an additional charged Higgs \cite{Crivellin:2015hha}.

\section{Measuring \CP violation in \Bs-\Bsb mixing}
Neutral mesons can oscillate into their own antiparticle through a second-order weak amplitude in a process called mixing. Since these processes are heavily suppressed in the Standard Model, they are sensitive to new physics. This analysis concerns the measurement of the \CP violation in mixing of \Bs mesons, which occurs when $\mathcal{P}(\Bs \to \Bsb) \neq \mathcal{P}(\Bsb \to \Bs)$. The amount of \CP violation in mixing can be measured using flavour-specific decays, of which the final state $f$ indicates whether the \B decayed as a \Bs or \Bsb. The flavour-specific asymmetry, $a_{\rm{fs}}$, is defined as:
\begin{equation}
a_{\rm{fs}} = \frac{\Gamma(\bar{B}_q \rightarrow B_q \rightarrow f) - \Gamma(B_q \rightarrow \bar{B}_q \rightarrow \bar{f})}{\Gamma(\bar{B}_q \rightarrow B_q \rightarrow f) + \Gamma(B_q \rightarrow \bar{B}_q \rightarrow \bar{f})}.
\label{eq:afs}
\end{equation}
For the \Bs mesons, this quantity is called \asls, and for the \Bd system, \asld. The Standard Model predictions for both are very small: $\asls = (2.22 \pm 0.27) \times 10^{-5}$ and $\asld = (-4.7 \pm 0.6) \times 10^{-4}$ \cite{Artuso:2015swg}. In their dimuon measurement, \dzero has found a discrepancy of a combination of \asls and \asld with respect to the SM value of 3$\sigma$ \cite{Abazov:2013uma}.

\begin{figure}[h]\centering
\includegraphics[width=0.43\textwidth]{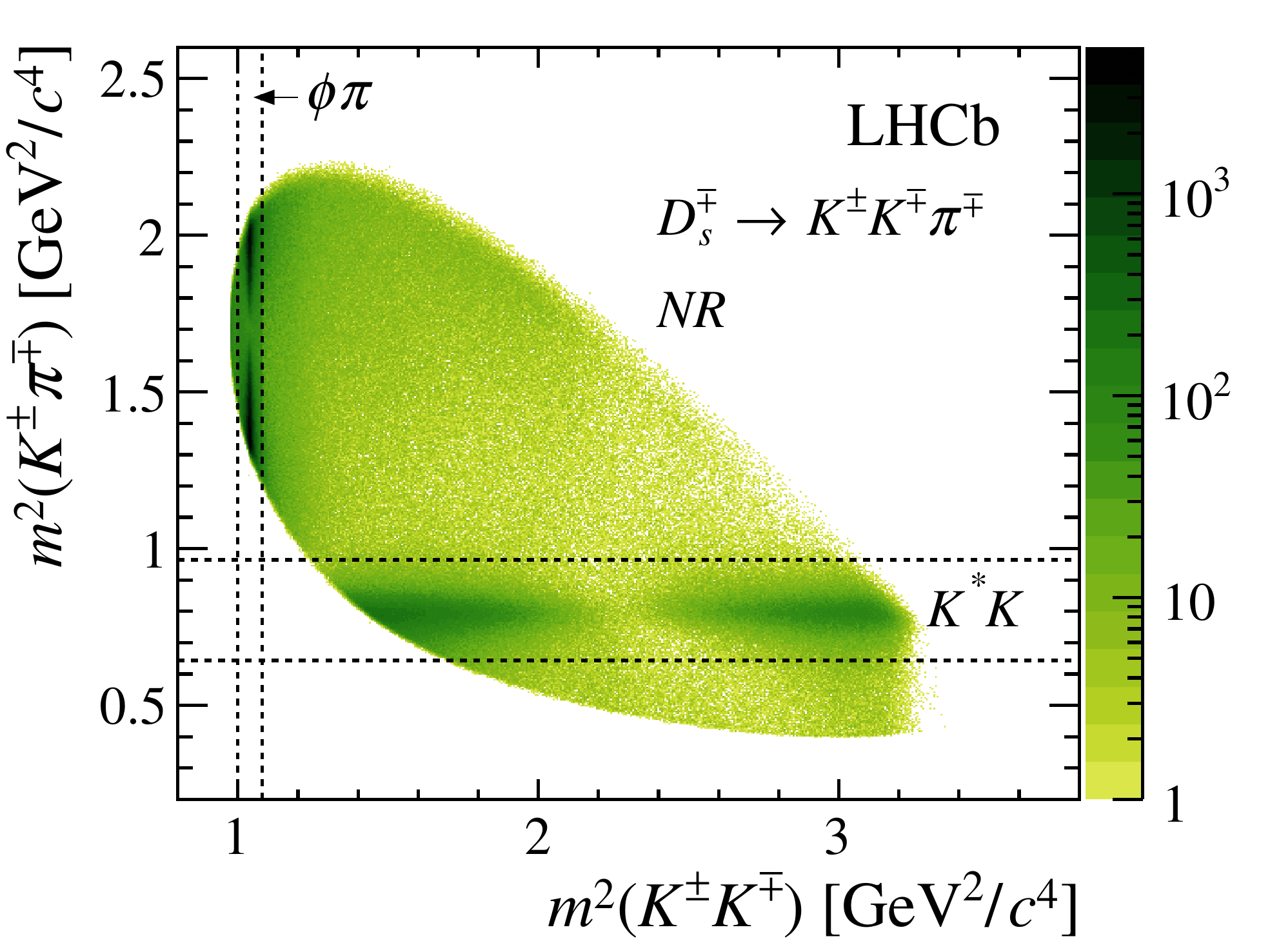}
\includegraphics[width=0.45\textwidth]{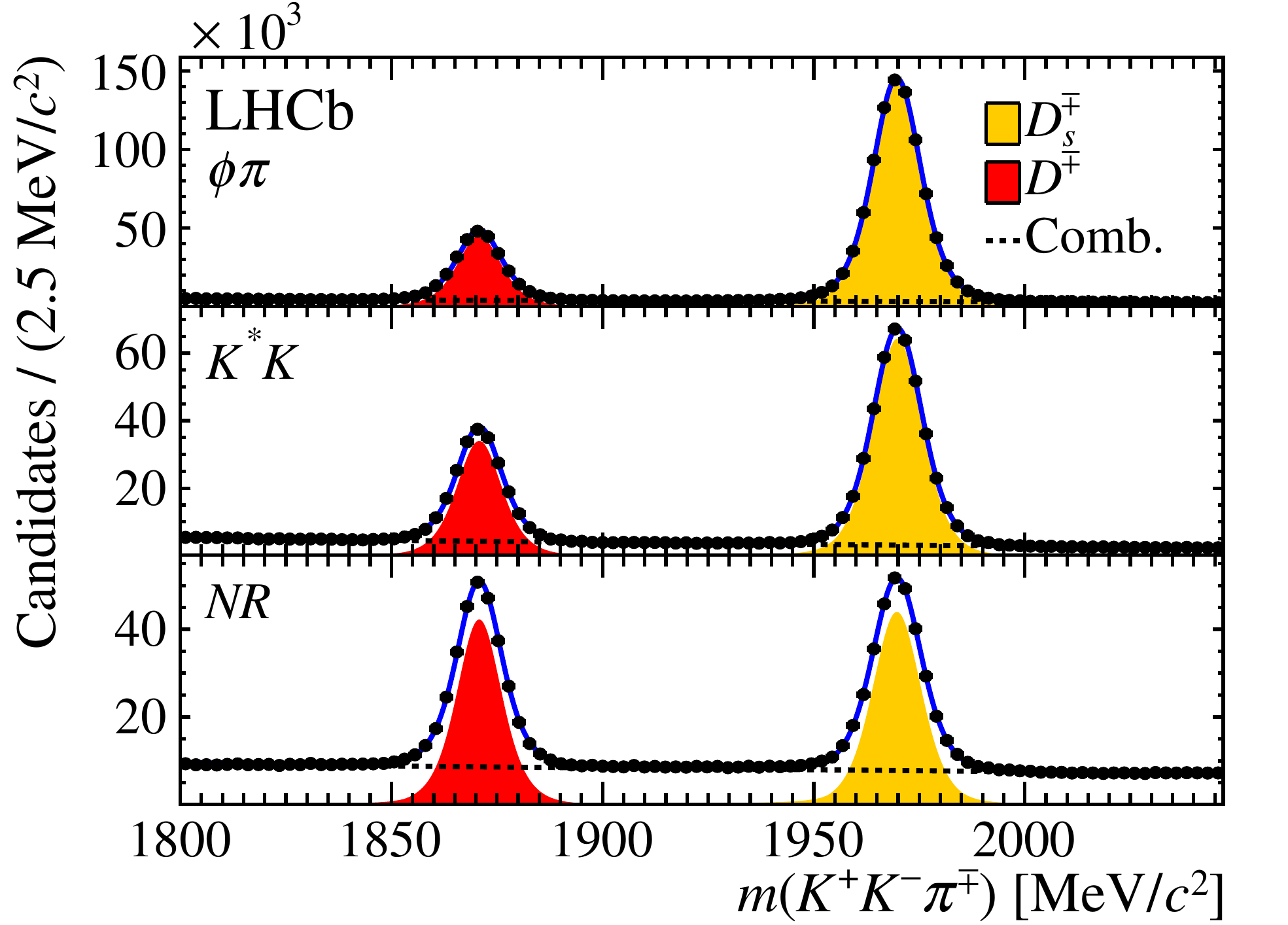}
\caption{\label{fig:asls_dalitz_mass} On the left: the Dalitz plot of the \Dsmp \to \Kpm \Kmp \pimp decay for the selected \Dsmp $\mu^{\pm}$ candidates, showing the three regions. On the right: the mass distributions of the \Kp \Km \pimp mass for the three Dalitz regions, overlaid with the results of the signal and combinatorial fit.}
\end{figure}

To measure \asls in \lhcb, the inclusive semileptonic decay $\Bs \to \Dsm \mup \neum$ is studied, where the \Dsm is reconstructed in the \Kp \Km \pimp final state. The full phase space of the \Dsm decay is analysed, divided into three regions of the Dalitz plane: the $\phi \pi$ region, where the \Kp \Km pair goes through the $\phi$ resonance; the $\Kstar \kaon$ region, where the \Kp \pim pair goes through the $\Kstar \rm{(892)}^0$ resonance; and the non-resonant region, NR. These regions are indicated in Fig.~\ref{fig:asls_dalitz_mass} (left). Backgrounds from partially reconstructed and misidentified decays are removed, as well as the backgrounds coming from \Bs from \D decays. The signal yields are obtained by matching \Dsm \mup pairs and fitting the \Dsm yields, which are shown in Fig.~\ref{fig:asls_dalitz_mass} (right).

\begin{figure}[h]\centering
\includegraphics[width=0.47\textwidth]{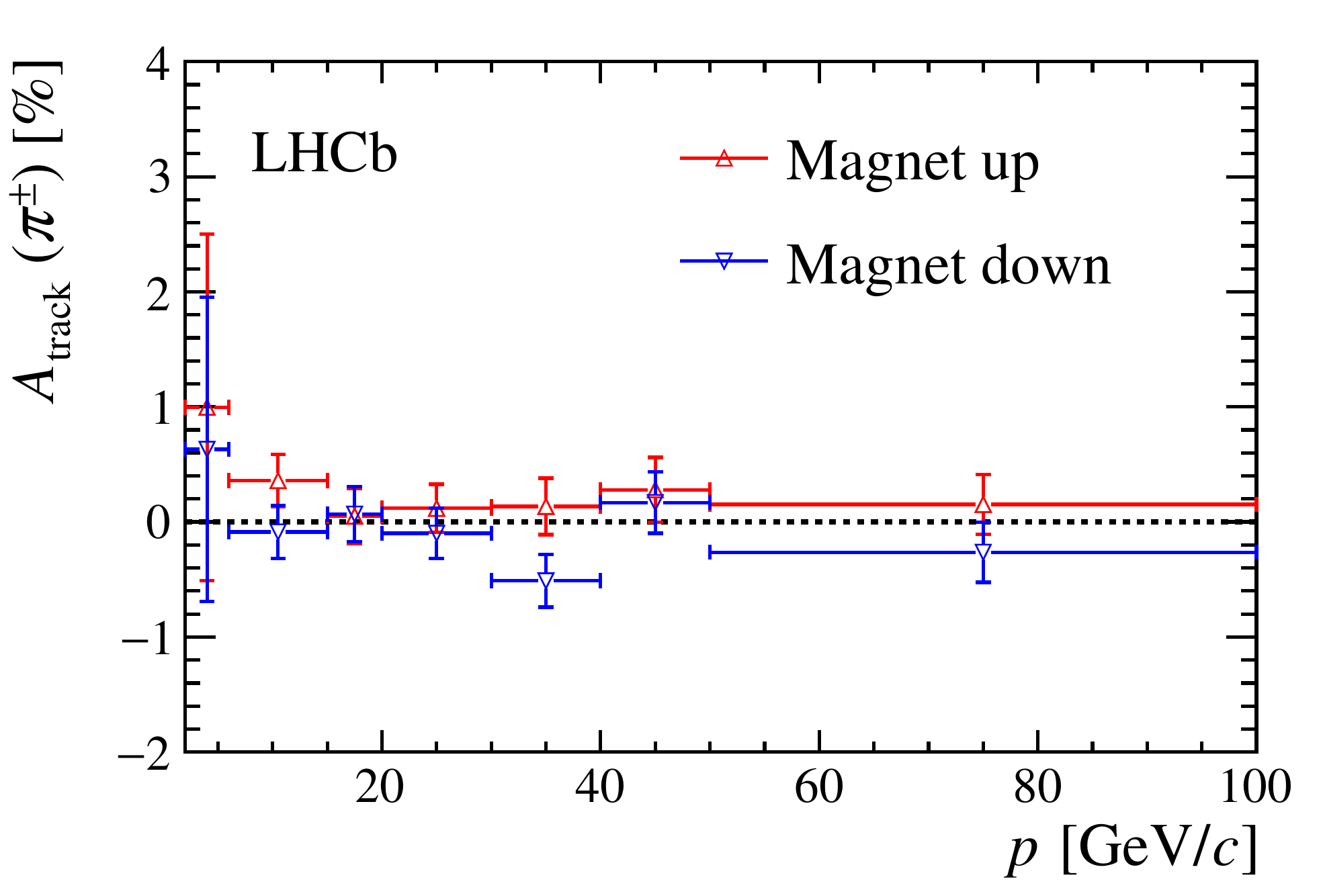}
\includegraphics[width=0.52\textwidth]{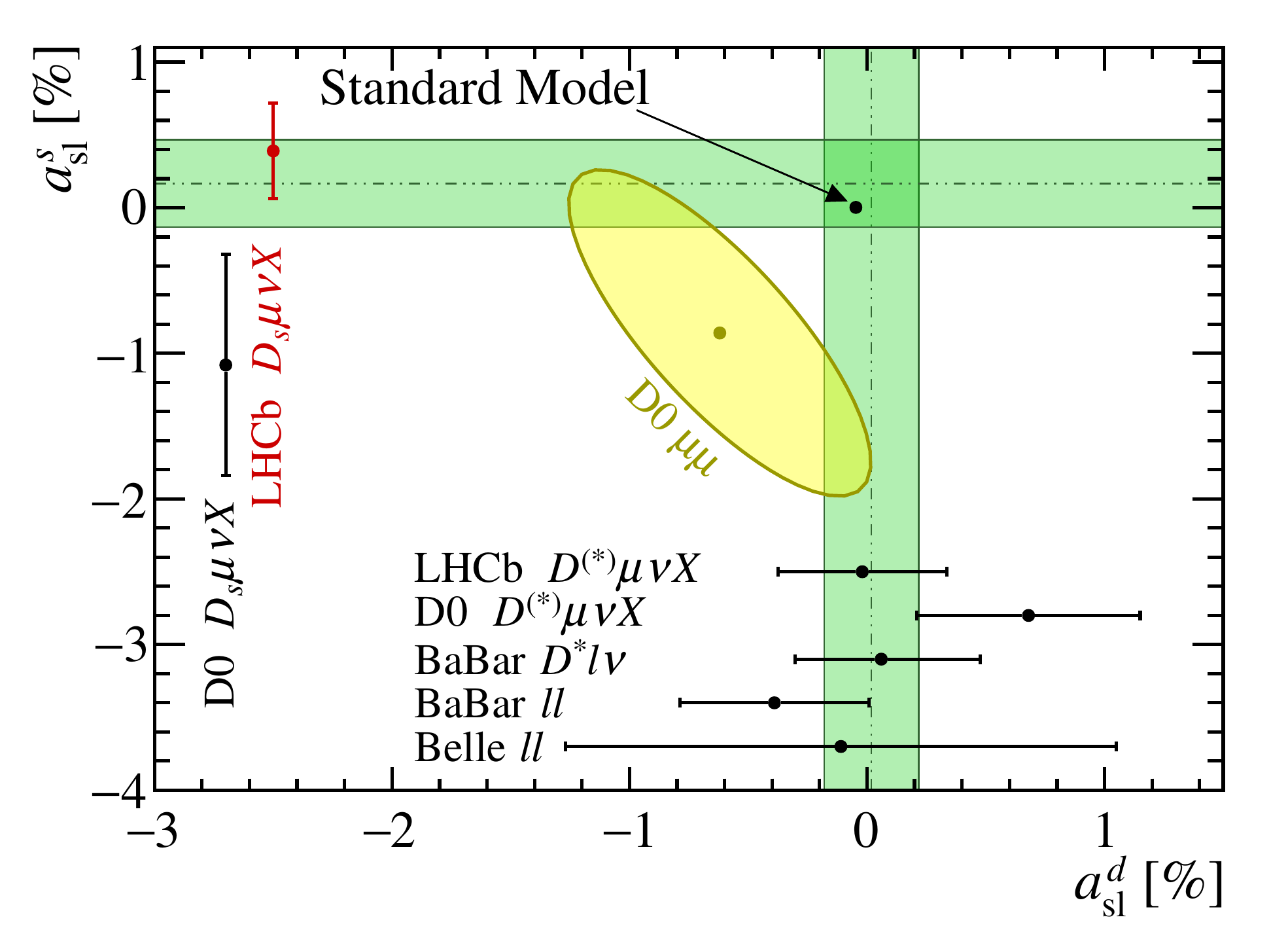}
\caption{\label{fig:asls_tracking_overview} On the left: the \pion tracking asymmetry as a function of momentum. On the right: overview of the current \asls and \asld measurements. The yellow ellipse shows the result of the \dzero dimuon measurement \cite{Abazov:2013uma}, while the points with error bars show the individual measurement of \asls or \asld from \lhcb \cite{LHCb-PAPER-2016-013, LHCb-PAPER-2014-053}, \babar \cite{Lees:2013sua, Lees:2014kep}, \belle \cite{Nakano:2005jb}, and \dzero \cite{Abazov:2012hha, Abazov:2012zz}. The green bands are the averages of \asls and \asld.}
\end{figure}

The asymmetry in the signal yields is the raw asymmetry, $A_{\rm{raw}}$, which is defined as:
\begin{equation}
A_{\rm{raw}} = \frac{N(D_s^- \mu^+) - N(D_s^+\mu^-)}{N(D_s^- \mu^+) + N(D_s^+\mu^-)}.
\label{eq:Araw}
\end{equation}
To measure \asls, this asymmetry needs to be corrected for the detection asymmetries, $A_{\rm{det}}$, and the asymmetries from peaking backgrounds. These are the production asymmetries from \B decays that peak in the \Dsm mass and called $A_{\rm{bkg}}$, while the fraction of these backgrounds is $f_{\rm{bkg}}$. \asls can then be determined as follows:
\begin{equation}
a_{\rm{sl}}^s=\frac{2}{1-f_{\rm{bkg}}}(A_{\rm{raw}}-A_{\rm{det}}-f_{\rm{bkg}}A_{\rm{bkg}}).
\label{eq:asls}
\end{equation}
The detection asymmetries arise from a difference in reconstruction efficiency between positively and negatively charged particles, and can be split in tracking, trigger and particle identification asymmetries. All of those are measured using data-driven methods. As the largest systematic in the previous measurement of \asls was that of the tracking asymmetry, the current analysis uses two different methods to determine these and the combination of these two is shown as a function of \pion momentum in Fig.~\ref{fig:asls_tracking_overview} (left).

The latest \lhcb measurement of \asls uses the full 3.0\invfb of the Run 1 dataset and measures $\asls = (0.39 \pm 0.26 \pm 0.20)\%$, where the first uncertainty is statistical and the second systematic \cite{LHCb-PAPER-2016-013}. The result shown in the overview plot in Fig.~\ref{fig:asls_tracking_overview} together with the measurements of \asls and \asld from \dzero \cite{Abazov:2012hha, Abazov:2012zz}, \babar \cite{Lees:2013sua, Lees:2014kep} and \belle \cite{Nakano:2005jb}. It is consistent with the Standard Model prediction and does not confirm the discrepancy arising from the \dzero dimuon analysis \cite{Abazov:2013uma}.

\vspace{-10pt}

\section{Conclusion}
This article presents two semileptonic measurements from the \lhcb collaboration. The \RDstar measurement provides a test of lepton universality and measures a value of $\RDstar = 0.336 \pm 0.027 \rm{(stat)} \pm 0.030 \rm{(syst)}$, which is 2.1$\sigma$ larger than the value expected from the Standard Model. The \asls analysis measures \CP violation in mixing of \Bs-\Bsb mesons and finds that $\asls = (0.39 \pm 0.26 \rm{(stat)} \pm 0.20 \rm{(syst)})\%$, which is the most precise measurement of \CP violation in the \Bs system to date and is consistent with the SM value.

\bibliographystyle{is-unsrt}
\bibliography{my-refs}

\begin{thebibliography}{10}
\ifx \showCODEN  \undefined \def \showCODEN #1{CODEN #1}  \fi
\ifx \showISBN   \undefined \def \showISBN  #1{ISBN #1}   \fi
\ifx \showISSN   \undefined \def \showISSN  #1{ISSN #1}   \fi
\ifx \showLCCN   \undefined \def \showLCCN  #1{LCCN #1}   \fi
\ifx \showPRICE  \undefined \def \showPRICE #1{#1}        \fi
\ifx \showURL    \undefined \def \showURL {URL }          \fi
\ifx \path       \undefined \input path.sty               \fi
\ifx \ifshowURL \undefined
     \newif \ifshowURL
     \showURLtrue
\fi

\bibitem{Alves:2008zz}
A.~A. Alves~Jr. et~al.
\newblock {The \lhcb detector at the LHC}.
\newblock {\em JINST}, 3:\penalty0 S08005, 2008.

\bibitem{LHCb-DP-2014-002}
R.~Aaij et~al.
\newblock {LHCb detector performance}.
\newblock {\em Int. J. Mod. Phys.}, A30:\penalty0 1530022, 2015.

\bibitem{Fajfer:2012vx}
S.~Fajfer, J.F. Kamenik, and I.~Nisandzic.
\newblock {On the $B \to D^* \tau \bar \nu_{\tau}$ Sensitivity to New Physics}.
\newblock {\em Phys. Rev.}, D85:\penalty0 094025, 2012.

\bibitem{LHCb-PAPER-2015-025}
R.~Aaij et~al.
\newblock {Measurement of the ratio of branching fractions ${\cal
  B}(\overline{B}^0 \to D^{*+} \tau^{-}\overline{\nu}_{\tau})/{\cal
  B}(\overline{B}^0 \to D^{*+} \mu^{-}\overline{\nu}_{\mu})$}.
\newblock {\em Phys. Rev. Lett.}, 115:\penalty0 111803, 2015.

\bibitem{Bozek:2010xy}
A.~Bozek et~al.
\newblock {Observation of $\Bp \to \Dstarzb \tau^+ \nu_\tau$ and Evidence for
  $\Bp \to \Dzb \tau^+ \nu_\tau$ at Belle}.
\newblock {\em Phys. Rev.}, D82:\penalty0 072005, 2010.

\bibitem{Huschle:2015rga}
M.~Huschle et~al.
\newblock {Measurement of the branching ratio of $\bar{B} \to D^{(\ast)} \tau^-
  \bar{\nu}_\tau$ relative to $\bar{B} \to D^{(\ast)} \ell^- \bar{\nu}_\ell$
  decays with hadronic tagging at Belle}.
\newblock {\em Phys. Rev.}, D92\penalty0 (7):\penalty0 072014, 2015.

\bibitem{Abdesselam:2016cgx}
A.~Abdesselam et~al.
\newblock {Measurement of the branching ratio of $\bar{B}^0 \rightarrow D^{*+}
  \tau^- \bar{\nu}_{\tau}$ relative to $\bar{B}^0 \rightarrow D^{*+} \ell^-
  \bar{\nu}_{\ell}$ decays with a semileptonic tagging method}.
\newblock 2016.

\bibitem{Lees:2012xj}
J.P. Lees et~al.
\newblock {Evidence for an excess of $\bar{B} \to D^{(*)} \tau^-\bar{\nu}_\tau$
  decays}.
\newblock {\em Phys. Rev. Lett.}, 109:\penalty0 101802, 2012.

\bibitem{Crivellin:2015hha}
A.~Crivellin, J.~Heeck, and P.~Stoffer.
\newblock {A perturbed lepton-specific two-Higgs-doublet model facing
  experimental hints for physics beyond the Standard Model}.
\newblock {\em Phys. Rev. Lett.}, 116\penalty0 (8):\penalty0 081801, 2016.

\bibitem{Na:2015kha}
H.~Na, C.M. Bouchard, G.P. Lepage, C.~Monahan, and J.~Shigemitsu.
\newblock {$B \rightarrow D l \nu$ form factors at nonzero recoil and
  extraction of $|V_{cb}|$}.
\newblock {\em Phys. Rev.}, D92\penalty0 (5):\penalty0 054510, 2015.
\newblock [Erratum: Phys. Rev.D93, no.11, 119906(2016)].

\bibitem{Lattice:2015rga}
J.A. Bailey et~al.
\newblock {\B \to D $\l \nu$ form factors at nonzero recoil and |V$_{cb}$| from
  2+1-flavor lattice QCD}.
\newblock {\em Phys. Rev.}, D92\penalty0 (3):\penalty0 034506, 2015.

\bibitem{Amhis:2014hma}
Y.~Amhis et~al.
\newblock {Averages of $b$-hadron, $c$-hadron, and $\tau$-lepton properties as
  of summer 2014}.
\newblock 2014.
\newblock
  {\href{http://www.slac.stanford.edu/xorg/hfag2/semi/summer16/summer16_dtaunu.html}{Link
  to plot}}.

\bibitem{Artuso:2015swg}
M.~Artuso, G.~Borissov, and A.~Lenz.
\newblock {CP Violation in the $B_s^0$ System}.
\newblock 2015.

\bibitem{Abazov:2013uma}
V.M. Abazov et~al.
\newblock {Study of \CP-violating charge asymmetries of single muons and
  like-sign dimuons in $p\antiproton$ collisions}.
\newblock {\em Phys.Rev.}, D89:\penalty0 012002, 2014.

\bibitem{LHCb-PAPER-2016-013}
R.~Aaij et~al.
\newblock {Measurement of the $C\!P$ asymmetry in $B^0_s-\overline{B}^0_s$
  mixing}.
\newblock {\em Phys. Rev. Lett.}, 117:\penalty0 061803, 2016.

\bibitem{Abazov:2012hha}
V.M. Abazov et~al.
\newblock {Measurement of the semileptonic charge asymmetry in $B^0$ meson
  mixing with the D0 detector}.
\newblock {\em Phys.Rev.}, D86:\penalty0 072009, 2012.

\bibitem{Abazov:2012zz}
V.M. Abazov et~al.
\newblock {Measurement of the semileptonic charge asymmetry using $B_s^0 \to
  D_s \mu X$ decays}.
\newblock {\em Phys.Rev.Lett.}, 110:\penalty0 011801, 2013.

\bibitem{Lees:2013sua}
J.~P. Lees et~al.
\newblock {Search for \CP violation in \Bd-\Bdb mixing using partial
  reconstruction of $\Bd \to D^{*-}X\ell^+ \nu_\ell$ and a kaon tag}.
\newblock {\em Phys.Rev.Lett.}, 111:\penalty0 101802, 2013.

\bibitem{Lees:2014kep}
J.~P. Lees et~al.
\newblock {Study of $CP$ Asymmetry in $B^0-\bar B^0$ Mixing with Inclusive
  Dilepton Events}.
\newblock {\em Phys. Rev. Lett.}, 114\penalty0 (8):\penalty0 081801, 2015.

\bibitem{Nakano:2005jb}
E.~Nakano et~al.
\newblock {Charge asymmetry of same-sign dileptons in \Bd-\Bdb mixing}.
\newblock {\em Phys.Rev.}, D73:\penalty0 112002, 2006.

\bibitem{LHCb-PAPER-2014-053}
R.~Aaij et~al.
\newblock {Measurement of the semileptonic $C\!P$ asymmetry in
  $B^0$--$\overline{B}^0$ mixing}.
\newblock {\em Phys. Rev. Lett.}, 114:\penalty0 041601, 2015.

\end{thebibliography}

\end{document}